\documentclass[10pt,conference]{IEEEtran}
\IEEEoverridecommandlockouts

\usepackage[table]{xcolor}
\usepackage{cite}
\usepackage{amsmath,amssymb,amsfonts}
\usepackage{algorithmic}
\usepackage{graphicx}
\usepackage{textcomp}
\usepackage{listings}
\usepackage{comment}
\usepackage{multirow}
\usepackage{booktabs}
\usepackage{listings}
\usepackage{tcolorbox}
\usepackage{makecell}
\usepackage[font=footnotesize,skip=8pt]{caption} %

\usepackage{hyperref}
\def\BibTeX{{\rm B\kern-.05em{\sc i\kern-.025em b}\kern-.08em
    T\kern-.1667em\lower.7ex\hbox{E}\kern-.125emX}}

\begin{document}

\title{Evaluating the Ability of GPT-4o to Generate Verifiable Specifications in VeriFast}
\author{
\IEEEauthorblockN{Wen Fan}
\IEEEauthorblockA{Purdue University \\
West Lafayette, IN, USA \\
fan372@purdue.edu}
\and
\IEEEauthorblockN{Marilyn Rego}
\IEEEauthorblockA{Purdue University \\
West Lafayette, IN, USA \\
mrego@purdue.edu}
\and
\IEEEauthorblockN{Xin Hu}
\IEEEauthorblockA{University of Michigan - Ann Arbor \\
Ann Arbor, MI, USA \\
hsinhu@umich.edu}
\and
\IEEEauthorblockN{Sanya Dod}
\IEEEauthorblockA{Purdue University \\
West Lafayette, IN, USA \\
sdod@purdue.edu}
\and
\IEEEauthorblockN{Zhaorui Ni}
\IEEEauthorblockA{Purdue University \\
West Lafayette, IN, USA \\
ni134@purdue.edu}
\and
\IEEEauthorblockN{Danning Xie}
\IEEEauthorblockA{Purdue University \\
West Lafayette, IN, USA \\
xie342@purdue.edu}
\and
\IEEEauthorblockN{Jenna DiVincenzo}
\IEEEauthorblockA{Purdue University \\
West Lafayette, IN, USA \\
jennad@purdue.edu}
\and
\IEEEauthorblockN{Lin Tan}
\IEEEauthorblockA{Purdue University \\
West Lafayette, IN, USA \\
lintan@purdue.edu}
}

\maketitle

\begin{abstract}
Static verification is a powerful method for enhancing software quality, but it demands significant human labor and resources. This is particularly true of static verifiers that reason about heap manipulating programs using an ownership logic. LLMs have shown promise in a number of software engineering activities, including code generation, test generation, proof generation for theorem provers, and specification generation for static verifiers. However, prior work has not explored how well LLMs can perform specification generation for specifications based in an ownership logic, such as separation logic.
To address this gap, this paper explores OpenAI’s GPT-4o model's effectiveness in generating specifications on C programs that are verifiable with VeriFast, a separation logic based static verifier. Our experiment employs three different types of user inputs as well as basic and Chain-of-Thought (CoT) prompting to assess GPT's capabilities. Our results indicate that the specifications generated by GPT-4o preserve functional behavior, but struggle to be verifiable. When the specifications are verifiable they contain redundancies. Future directions are discussed to improve the performance. 
\end{abstract}


\section{Introduction}
Auto-active (Hoare-logic styled \cite{hoare1969axiomatic}, static) verifiers, such as Viper \cite{muller2016viper}, Verus \cite{lattuada2023verus}, Dafny \cite{leino2010dafny}, Gillian \cite{fragoso2020gillian}, and VeriFast \cite{jacobs2011verifast},  are powerful as they can prove the absence of large classes of bugs in code. Ideally, users of such tools need only specify the intended behavior of their code on the code itself (as pre- and postconditions), and the tool will automatically provide feedback on whether or not the code is provably correct with respect to this behavior. In reality, auto-active verifiers require many more auxiliary specifications (e.g. loop invariants, lemmas, opens, closes) to achieve this goal, burdening users.

In recent years, large language models (LLMs) have been effective in generating code \cite{chen2022codet,sarsa2022automatic}, test-cases \cite{deng2023large,lemieux2023codamosa,rao2023cat,schafer2023empirical,wang2024software,xia2024fuzz4all}, and proofs in proof assistants \cite{zheng2023lyra,yang2024leandojo, jiang2021lisa,welleck2023llmstep,first2023baldur}. LLMs have also been shown to be effective at generating specifications supported by auto-active verifiers \cite{ma2024specgen,kamath2023finding,misu2024towards,he2024beyond,mugnier2024laurelgeneratingdafnyassertions,mukherjee2024automatedverificationllmsynthesizedc}. However, related work has not explored whether or not off-the-shelf LLMs can generate specifications based on a permissions logic, like \textit{separation logic} \cite{reynolds2002separation}, that can be verified by auto-active verifiers such as VeriFast, Gillian, and Viper. 
Thanks to such specifications, these verifiers do well at verifying programs that manipulate the heap for both memory safety and functional properties. But, permissions logic based specifications are particularly cumbersome to write, because they must specify the shape of the heap alongside functional constraints. This leads to specifications containing a number of predicates that hide heap details; and as a result, numerous lemmas, folds, unfolds, and special loop invariants that are used to connect the content of these predicates. While such specifications are difficult to reason about, they are written in a patterned way that may be amenable to generation via LLMs. 

Therefore, this paper evaluates how effective OpenAI's GPT-4o model \cite{openai2024gpt4o} is at generating specifications that can be verified by VeriFast \cite{jacobs2011verifast}, which supports separation logic-based verification of C and Java code. We selected OpenAI's GPT-4o model due to its strong text comprehension and generation capabilities \cite{achiam2023gpt}. We ask three main research questions regarding whether or not the LLM's output specifications and code preserve functional behavior, are verified, and are conventional. To answer these questions, we developed 21 input-output pairs from 150 publicly available VeriFast programs as test cases and ground truth, and employed two prompt engineering methods (Basic Prompting and CoT Prompting) to instruct GPT-4o. Then, we manually inspected the outputs generated by the LLM by comparing them with the ground truth in a qualitative analysis. Results show that GPT-4o generates specifications and code that preserves functional behavior expressed in input files. We also found that both prompting methods result in tons of verification errors, and the small number of LLM output files that do verify contain redundant specifications. This shows a need for better prompting techniques, which we intend to explore in follow-up work.

\section{Related Work}
\subsection{Specification Generation for Auto-active Verifiers}

Houdini \cite{flanagan2001houdini} is an annotation assistant that suggests candidate annotations using heuristics, which are then verified or refuted by ESC/Java \cite{leino2000esc}. %
Vogels et al. \cite{vogels2011annotation} showed on eight small programs that given preconditions shape analysis can deduce all specifications required to verify programs with VeriFast. %
Yao et al. \cite{yao2023leveraging} combine GPT-4 and static analysis to automate proof synthesis in Verus \cite{lattuada2023verus}. Kamath et al. \cite{kamath2023finding} use LLMs to generate inductive loop invariants and then verify the candidates with Frama-C \cite{kirchner2015frama}. Misu et al. \cite{misu2024towards} investigated how well GPT-4 and PaLM-2 can synthesize verified methods in Dafny with various prompting approaches, and found few-shot reasoning is the best.
Laurel \cite{mugnier2024laurel} uses large language models with prompting techniques to generate helper assertions for Dafny programs. Finally, Mukherjee et al. \cite{mukherjee2024automatedverificationllmsynthesizedc}, which is most closely related, present a synthesis framework for C code that uses LLMs to generate programs verified by VeriFast. Our work focuses on verifying existing C code with VeriFast rather than generating whole programs. Further, our work considers loops and helper functions; theirs does not.

\subsection{Program Invariant Generation with LLMs or ML}
Daikon \cite{ernst2007daikon} automatically infers likely invariants from program executions using machine learning for use in testing, debugging, and verification tasks. Pei et al. \cite{pei2023can} used fine-tuned large language models to predict program invariants on par with Daikon.
Xie et al. \cite{xie2023impact} evaluated the capabilities of LLMs to generate specifications from comments and documentation, demonstrating few-shot learning and advanced prompt strategies outperform traditional methods by 5.1–10.0\%.
Only Nimmer et. al \cite{nimmer2002automatic} tried to integrate generated program invariants with an auto-active verifier pairing Daikon with ESC/Java to achieve over 90\% precision and recall. Daikon does not rely on LLMs to generate specifications and ESC/Java does not support separation logic, so our work is novel in comparison.

\subsection{Proof Synthesis for Proof Assistants with LLMs or ML}

Lots of work utilize LLMs or neural networks to generate tactics (next proof steps) or whole proofs in proof assistants (interactive theorem provers) such as Coq \cite{barras1997coq}, Isabelle/HOL \cite{nipkow2002isabelle}, and Lean \cite{de2015lean}.
For Coq, Sanchez-Stern et al. \cite{sanchez-stern2023passport} improved on existing neural network based proof synthesis tools ASTactic \cite{yang2019learning} and TacTok \cite{first2020tactok} by modeling identifiers. Lu et. al \cite{lu2024proof}  combined LLMs with symbolic methods to synthesize whole proofs.
For Isabelle/HOL, Jiang et. al. \cite{jiang2021lisa} use a fine-tuned language model to synthesize proof steps until a proof is achieved. %
Thor \cite{jiang2022thor} uses hammers for premise selection and designates the rest of proof synthesis to language models. %
Similarly, Jiang et al. \cite{jiang2023draft} leverage informal and formal proof sketches that are filled in by LLMs. %
Finally, First et. al \cite{first2023baldur} fine-tune LLMs for whole-proof generation and repair. %
For Lean, Han et al. \cite{han2022proof} use LLMs to suggest proof tactics. %
In contrast, our work focuses on the capabilities of LLMs to generate specifications for  auto-active verifiers.

\section{VeriFast \& Benchmarks}

\subsection{VeriFast}

\begin{table*}[h]
\centering
\renewcommand{\arraystretch}{1.2} %
\begin{tabular}{|c|c|c|c|c|c|c|}
\hline
\textbf{Specification} & Pre. & Post. & Open & Close & Pred. Declar. & Lemma Declar. \\ \hline
\textbf{Aggregated Result} & 8/3/59/35/6/3 & 9/2/47/35/5/5 & 0/6/0/33/0/0 & 0/3/0/38/0/0 & 24/5/15/5/2/10 & 0/0/9/0/1/0 \\ \hline
\end{tabular}
\caption{Aggregated results of specification constructs. Element counts are formatted as ``Separation Logic Arrow/Fractional Permissions/Boolean Expression/Predicate Call/Empty Heap/Malloc Block."}
\label{tab:aggregated-results}
\end{table*}

VeriFast \cite{jacobs2011verifast} is a sound and modular, auto-active verifier that reasons with symbolic execution \cite{king1976symbolic} to efficiently verify single and multi-threaded C and Java programs against specifications in separation logic \cite{reynolds2002separation}. As a result, VeriFast can verify functional properties and memory safety (e.g., dangling pointer dereference and double free).
We chose VeriFast due to its maturity: e.g. verifying four industrial case studies for memory safety \cite{philippaerts2014software}. Furthermore, over 150 open-source C programs have been specified and verified with VeriFast. %

\subsection{VeriFast Benchmarks used in the Evaluation}
\label{sec:benchmarks}

We selected a subset of 21 fully specified C files available publicly in \href{https://github.com/verifast/verifast/tree/c7a1817a550005d3ae830d15df06ccaaf840da49/examples}{VeriFast's Github repository} to use in our evaluation. 
These benchmarks cover diverse programming concepts and verification properties. For example, \textit{filter\_stack.c} and \textit{values.c} involve on-heap data structures like stacks and linked lists, while others have file I/O (e.g., \textit{cp.c}), fractional permission (e.g., \textit{fractions-counting.c}) and recursion (e.g., \textit{wc.c}).
We extracted the types of specification constructs used in each file (e.g. separation logic arrow and predicate instance) and where they are specified (e.g., in precondition or postcondition). Table \ref{tab:aggregated-results} presents the aggregated results of files from this analysis. Of the 398 specification elements observed, predicate calls (44.2\%), boolean expressions (32.7\%) and separation logic arrows (10.3\%) are prevalent. Moreover, the heap-related specifications (e.g., separation logic arrow and fractional permission) occurred the most in preconditons, postconditions and predicate declarations. We also find that the number of specifications varied a lot (e.g. \textit{filter\_stack.c} has 49 lines while \textit{typedef.c} has 2 lines).

\subsection{Input-Output Pairs}
\label{sec:io-pairs}

We developed input-output pairs from the benchmarks described in \S\ref{sec:benchmarks}. The inputs are GPT-4o's starting point and the outputs are GPT-4o's goal. Thus, the output files are defined as the fully verified benchmarks. %
We experiment with three types of inputs per distinct output file that represent possible user inputs: Natural Language (NL), Functional Behavior (FB), and Functional Behavior Plus (FB+). Each of the input types contain code from the output file but with different partial specifications. An NL input does not contain any formal specifications, but only a natural language description of the intended behavior of each function.
A FB input contains formal specifications that specify only the functional behavior of each function (e.g., pre- and postconditions). %
Finally, a FB+ input contains only formally specified preconditions and postconditions directly from the output file, which specify functional behavior but may also contain other properties.
Examples of each input type and the corresponding output for them is shown in Listing \ref{lst:IO-pairs-eg}.

\begin{lstlisting} [caption={Input-output pairs created for the \texttt{increment} function}, label={lst:IO-pairs-eg}]
// The increment function increments the 
// value of the Counter structure by one ...
void increment(struct Counter* c) // NL input
{ int tmp = c->value; c->value = tmp + 1; }

void increment(struct Counter* c) // FB input
//@ requires Counter(c, ?v);
//@ ensures Counter(c, v+1);
{ int tmp = c->value; c->value = tmp + 1; }

void increment(struct Counter* c) // FB+ input
//@ requires Counter(c, ?v) &*& v < INT_MAX;
//@ ensures Counter(c, v+1);
{ int tmp = c->value; c->value = tmp + 1; }

void increment(struct Counter* c) // output
//@ requires Counter(c, ?v) &*& v < INT_MAX;
//@ ensures Counter(c, v+1);
{ //@ open Counter(c, v);
  int tmp = c->value;
  c->value = tmp + 1;
  //@ close Counter(c, v+1); }
\end{lstlisting}

\section{Prompt Engineering}
\label{sec:prompting}

We explore two different prompting approaches in our study: Basic and Chain of Thought (CoT) prompting. %

\paragraph*{Basic Prompting}
The basic prompting approach instructs GPT-4o to generate verifiable specifications with VeriFast for an input with a single, few sentence prompt. %
This prompt provides minimal context and instructions and is available (\textit{link available post review}). %
Basic prompting establishes a baseline to evaluate CoT and future prompting approaches against.  

\paragraph*{CoT Prompting}
The CoT prompting approach \cite{wei2022chain} guides an LLM through structured, step-by-step instructions for generating VeriFast specifications. Our version of CoT mirrors the logical progression followed by experts in specification writing. For example, when writing a precondition, the prompt first explains how to capture the input behavior of a function, then details constraints on syntax and positional requirements, and finally addresses additional properties such as memory safety. The CoT prompt, with code, is available (\textit{Link included post review}). Unlike few-shot prompting used by Misu et al. \cite{misu2024towards}, our CoT prompting deconstructs the specification writing process into actionable steps.%

\section{Qualitative Analysis}
We prompt GPT-4o for output files using scripts implementing the prompting approaches (\S\ref{sec:prompting}) and our input files (\S\ref{sec:io-pairs}). %
We assess the GPT-4o outputs using a qualitative analysis designed to answer the following research questions: %
\begin{itemize}
    \item \textbf{RQ1:} How well does GPT-4o preserve functional behavior?
    \item \textbf{RQ2:} How well does GPT-4o generate verifiable specifications?
    \item \textbf{RQ3:} How conventional are the verifiable specifications generated by GPT-4o?
\end{itemize}
For each RQ, we developed an initial set of qualitative codes from our intuition, comparisons with the ground truth, and a pilot study on a subset of the output files. As we performed our qualitative analysis by applying the codes to parts of the output files as necessary, we refined our codes as well.

For \textbf{RQ1}, we assessed how well GPT-4o preserves functional behavior expressed in function contracts (precondition/postconditions) and in source code. %
We assigned codes/sub-codes categorized as \textit{preserved} when the output is equivalent to the input, \textit{strengthened} when the output implies the input, \textit{weakened} when the output is implied by the input, and \textit{others} (e.g. when the output is unrelated to the input).

For \textbf{RQ2}, we ran VeriFast on all outputs from GPT-4o to check their verifiability.
If a file was verified, then we analyzed its specifications for how conventional they are (\textbf{RQ3}). Otherwise, we assigned codes representing the cause of verification failure(s). To capture all the underlying verifiability issues in the output, we iteratively fixed failures until the output verifies.
The codes fall into two categories: \textit{compilation error} or \textit{verification error}. %
Compilation error codes include \textit{specification out-of-position (spec-OOP)}, \textit{syntax errors} (errors during the parsing stage) and \textit{include \& type check errors} (errors during the include or type checking stage).
Verification errors %
are assigned based on the the component of the corresponding fix and include
incorrect \textit{precondition/postcondition}, \textit{predicate definition}, \textit{open/close/assert/leak} use or definition, \textit{lemma definition}, \textit{lemma use}, and \textit{loop invariant} and \textit{other} errors (e.g., source code being modified incorrectly). %

For \textbf{RQ3}, we analyze the conventionality of correctly generated specifications. Our codes for this research question capture redundant specifications that are unnecessary for verification and ambiguous specifications that exhibit naming mismatches with the intended function. %
The assigned codes are aggregated by occurrence; and, the results given next (\S\ref{sec:results}).

\section{Result \& Discussion}
\label{sec:results}

\subsection{\textbf{RQ1} - Functional Behavior Analysis}
Table \ref{tab:fb-code-result} summarizes functional behavior in output files, where most of them preserve the functional behavior. Specifically, in the 126 output files analyzed (21 benchmarks $\times$ 3 input types $\times$ 2 prompting), only 20 show changes in preconditions/postconditions and 3 show changes in source code. For the 7 files with strengthened functional behavior, 6 had extra constraints in postcondition (e.g., ``\texttt{new\_count <= count}"). For 10 files with weakened functional behavior, 9 missed properties in the postconditions or predicates (e.g., not specifying the value of balance in ``\texttt{acc->balance |-> ?b}''). The modified functional behavior in the source code was due to LLM altering the source code (e.g., changing the argument from -100 to 100). Thus, to answer \textbf{RQ1}, we show GPT-4o largely preserves functional behavior in specifications and source code across all input types and prompting techniques. %

\begin{table}
\centering
\begin{tabular}{|l|c|c|}
\hline
\textbf{Functional Behavior}       & Pre/Postcondition & Source Code \\ \hline
preserved     & 106                        & 123               \\ \hline
strengthened  & 7                          & 0                 \\ \hline
weakened      & 10                          & 1                 \\ \hline
others        & 3                          & 2                 \\ \hline
\textbf{Total}& 126                        & 126               \\ \hline
\end{tabular}
\caption{Functional Behavior Preservation Analysis}
\label{tab:fb-code-result}
\end{table}

\subsection{\textbf{RQ2} - VeriFast Specifications Analysis}
Table \ref{tab:error-code-result} shows the aggregated number of different errors in GPT-4o's outputs across the 21 benchmarks with different prompting methods and input types. Only 9 out of the 126 output files were directly verifiable, while the remaining files showcased a high number of errors, with 539 errors from basic prompting and 555 from the CoT prompting approach.

\subsubsection{Impact of Input Type}
The NL inputs resulted in more compilation errors (103 and 123) than other inputs (at most 28). Similarly, considering the errors of pre/postcondition and predicate, FB+ inputs resulted in fewer such errors (at most 6 + 2 = 8 in CoT prompt of FB+), compared to other types (at least 15 + 9 = 24). This difference can be attributed to more detailed and precise specifications provided in FB+ inputs. Thus, providing GPT-4o with examples or input containing detailed preconditions and postconditions in VeriFast syntax can improve the correctness of its output. However, even if correct preconditions/postconditions are provided, GPT-4o occasionally modifies them, introducing errors (e.g., removing bound checks in a precondition). Additionally, the three input types produced a similar number of errors for lemmas and loop invariants. For lemmas, this may be because they often require extra information beyond the capabilities of the GPT-4o. For instance, 114 errors were about implicit bound checks for data types like \texttt{size\_t}, and 27 errors were linked to meeting the preconditions of standard library functions such as \texttt{fread}. For loop invariants, the difficulty likely arises from the complexity of ensuring their correctness, as they must hold true both at the start of the loop and after each iteration. %

\subsubsection{Impact of Prompting Method}
The basic prompting and CoT prompting resulted in similar performance. On NL, FB, and FB+ inputs, the CoT prompt did not reduce the number of compilation errors compared to basic prompting.
In fact, for NL inputs, the CoT prompting resulted in more \textit{spec-OOP} errors (72 compared to 54). %
This is surprising because the CoT prompt explicitly told GPT-4o to put preconditions and postconditions at the right position.
Similarly, CoT is not significantly better than basic prompting on verification errors, except errors about open/close/assert/leak statements on FB+ input (47 compared to 77). %
This shows that in almost all cases adding more instructions steps in a prompt does not improve GPT-4o's performance. However, when preconditions, postconditions, and their predicate dependencies are specified as needed for verification, CoT works well for generating dependent auxiliary specifications. 
Thus, to answer \textbf{RQ2}, GPT-4o generates VeriFast specifications with limited success across the input types and prompting strategies assessed. Improvements to the prompting approaches or new prompting approaches should be explored.

\begin{table}
\centering
\footnotesize %
\setlength{\tabcolsep}{4pt} %
\renewcommand{\arraystretch}{1.2} %
\begin{tabular}{p{3cm}cccccc}
\toprule
\multicolumn{1}{c}{\textbf{Error Code}} & \multicolumn{3}{c}{\textbf{Basic Prompt}} & \multicolumn{3}{c}{\textbf{CoT Prompt}} \\ 
\cmidrule(lr){2-4} \cmidrule(lr){5-7}
\textbf{Sub-code} & \textbf{NL} & \textbf{FB} & \textbf{FB+} & \textbf{NL} & \textbf{FB} & \textbf{FB+} \\ 
\midrule
\multicolumn{7}{l}{\textbf{Compilation Error}} \\
spec-OOP & 54 & 1 & 0 & 72 & 8 & 0 \\ 
syntax & 16 & 10 & 5 & 12 & 7 & 9 \\ 
include \& type check & 33 & 8 & 9 & 39 & 13 & 10 \\ 
\textbf{Total} & \textbf{103} & \textbf{19} & \textbf{14} & \textbf{123} & \textbf{28} & \textbf{19} \\ 
\midrule
\multicolumn{7}{l}{\textbf{Verification Error}} \\
pre/postcondition & 28 & 18 & 2 & 25 & 15 & 6 \\ 
predicate definition & 10 & 12 & 3 & 14 & 9 & 2 \\ 
open/close/assert/leak & 49 & 65 & 77 & 57 & 65 & 47 \\ 
lemma definition & 1 & 2 & 2 & 2 & 2 & 5 \\ 
lemma use & 24 & 28 & 28 & 27 & 27 & 27 \\ 
loop invariant & 9 & 10 & 9 & 9 & 10 & 12 \\ 
others & 8 & 7 & 11 & 9 & 8 & 7 \\ 
\textbf{Total} & \textbf{129} & \textbf{142} & \textbf{132} & \textbf{143} & \textbf{136} & \textbf{106} \\ 
\bottomrule
\end{tabular}
\caption{Error Code Analysis for Basic and CoT Prompts}
\label{tab:error-code-result}
\end{table}

\subsection{\textbf{RQ3} - Convention Analysis}
In the verifiable output, there were 10 cases of redundant specifications that, while verifiable, could be simplified. This includes 4 specifications with unnecessary encapsulation of a predicate or lemma, 5 with redundant conditions, and 1 with an unused predicate. For example, in the \texttt{person} predicate definition in Listing \ref{lst:redundant-eg}, removing \texttt{p != 0} doesn't affect correctness since other clauses imply it.

\begin{lstlisting} [caption={Predicate with a redundant condition}, label={lst:redundant-eg}]
predicate person(struct person *p, ...) =
    p != 0 &*& malloc_block_person(p) &*& ...
\end{lstlisting}

Thus, to answer \textbf{RQ3}, GPT-4o-generated verifiable specifications occasionally include redundant elements, suggesting the need to make GPT-4o aware of conventions.

\subsection{Discussion and Suggestions for Future Work}
Future work can resolve errors by a more fine-grained prompt engineering method, such as refining prompt or input granularity (e.g., granularity of function as in \cite{misu2024towards}\cite{yao2023leveraging}\cite{kamath2023finding}), or providing verifier's feedback to LLM to guide the fix (e.g., \cite{kamath2023finding}). Further, the performance differences among input types show that providing example specifications to an LLM as part of the prompting process may improve performance. Few-shot prompting showed promise in related work \cite{misu2024towards}\cite{mugnier2024laurelgeneratingdafnyassertions}, and we plan to evaluate this approach in future work. Fine-tuning (e.g., \cite{chen2024automated}) may also be an interesting avenue for exploration.

\paragraph*{Threats to Validity}
The results of our analysis is limited with only 21 benchmarks, 1 verifier, and 1 LLM, but provides useful results to build on. %
The benchmarks were open-sourced years ago, so GPT-4o may have been trained on them biasing our results; however, GPT-4o's poor performance signals that this bias is not impactful.
Finally, while the analysis is methodical, it remains subjective; efforts to mitigate bias include thorough review, discussion, and consensus among conductors.

\section{Conclusion}
This work is the first attempt to evaluate LLMs' ability to generate separation logic specifications for verification with an auto-active verifier. We develop 21 benchmarks and design three types of user input and two prompting methods to prompt GPT-4o with. The results show that while the output generated by GPT-4o preserves functional behavior in the input, it suffers from significant errors that block automated verification. Despite more negative results, our work provides guidance for the development of more effective prompt engineering approaches.

\bibliographystyle{./IEEEtran}
\bibliography{./reference}

\end{document}